\documentclass[amsmath,amssymb,aps,floatfix,jcp,superscriptaddress,preprint]{revtex4-2}
\usepackage[utf8]{inputenc} 
\usepackage[T1]{fontenc}    
\usepackage{amsfonts}       
\usepackage{booktabs}       
\usepackage{hyperref}       
\usepackage{url}            
\usepackage{microtype}      
\usepackage{nicefrac}       
\usepackage{paralist}       
\usepackage{graphicx}
\usepackage{float}
\usepackage{wrapfig}
\usepackage[dvipsnames]{xcolor}
\usepackage{ulem}
\usepackage{tikz}
\usetikzlibrary{arrows,shapes,angles,quotes}
\usetikzlibrary{decorations.pathreplacing}
\usepackage{fullpage}

\usepackage{lmodern}
\usepackage{ulem}

\providecommand{\keywords}[1]
{
  \small	
  \textbf{\textit{Keywords---}} #1
}

\begin{document}
	
\title{Crystal Nucleation in Al-Ni Alloys: \\an Unsupervised Chemical and Topological Learning Approach}
\author{S\'ebastien Becker}
\affiliation{Univ. Grenoble Alpes, CNRS, Grenoble INP, SIMaP, F-38000 Grenoble, France}
\affiliation{Univ. Grenoble Alpes, CNRS, Grenoble INP, LIG, F-38000 Grenoble, France}
\author{Emilie Devijver}
\affiliation{CNRS, Univ. Grenoble Alpes, Grenoble INP, LIG, F-38000 Grenoble, France}
\author{R\'emi Molinier}
\affiliation{Univ. Grenoble Alpes, CNRS, IF, F-38000 Grenoble,France}
\author{No\"{e}l Jakse}
\affiliation{Univ. Grenoble Alpes, CNRS, Grenoble INP, SIMaP, F-38000 Grenoble, France}

\date{\today }


\begin{abstract}
Crystallization represents a fundamental process engendering solidification of a material and determines its microstructure. Driven by complex phenomena at the atomic scale, its understanding for alloys still remains elusive. The present work proposes a large scale molecular dynamics simulation study of the homogeneous crystal nucleation pathways of prototypical undercooled Al-Ni binary alloys. An unsupervised topological learning analysis shows that the nucleation sets in  first from a chemical ordering, followed by a bond-orientational ordering of the underlying crystal phase. Our results indicate also a different polymorph selection that depends on composition. While the nucleation pathway of Al\textsubscript{50}Ni\textsubscript{50} displays a single step with the emergence of B2 short-range order, a step-wise nucleation toward the L1\textsubscript{2} phase is seen for Al\textsubscript{25}Ni\textsubscript{75}. The influence of the nucleation of pure Al and Ni counterparts is further discussed.
\end{abstract}
\maketitle
\newpage

\section{Introduction}
\label{Sec:Introduction}

Understanding homogeneous crystal nucleation during which an undercooled liquid morphs into its underlying crystalline phase is of importance from a fundamental point of view as well as on the application side for materials manufacturing in industrial applications \cite{Kelton2010,Sosso2016}. Its intimate complex mechanisms take their roots at the atomic level and involve local symmetry breaking that can hardly be observed experimentally until very recently for Fe-Pt binary metallic nanoparticles by atomic electron tomography \cite{Zhou2019}. However, especially for metallic materials, identifying the early stages of nucleation from an experimental point of view is still merely out of reach, and simulations at the atomic scale remain largely the dedicated tools, applied to generic models \cite{Auer2001,ten1995,Kawasaki2010,Toxvaerd2020}, pure metals \cite{Shibuta2017,An2018,Mahata2018,Louzguine2020,Becker2020,Becker2022} and alloys \cite{Watson2011,Tang2013,Ko2017,Lin2018,Song2018,Tang2018,Sun2019,Orihara2020} to name a few.  

In classical nucleation theory, crystal nucleation and its rate can be attributed essentially to two thermodynamic factors, namely, the enthalpy difference between the crystal and liquid at the melting point, and effective crystal-liquid interfacial free energy. They depend also on atomic transport in the liquid state such as the diffusion \cite{Kelton2010,Sosso2016}.  The seminal work of Turnbull \cite{Turnbull1952} on the ability to deeply undercool monatomic liquid metals led to consider that the local atomic ordering of the undercooled melt might be incompatible with the crystalline structure, thus impeding crystal nucleation. It was later shown by Frank \cite{Frank1952} that the icosahedral ordering is locally more stable than the crystalline counterpart, which has triggered many theoretical and experimental works \cite{Reichert2000,Royall2015} in terms of the variety of polymorphs \cite{Ronceray2011}, competing short-range orders \cite{Jakse2003,Becker2020,Pedersen2021}, or an interplay between chemical and five-fold symmetry orderings in the case of liquid alloys \cite{Jakse2008,Pasturel2017,Tang2018}. For the latter multi-component systems, this raises naturally the question of the nature of the chemical and topological local orderings, and how they play a role at the onset of nucleation, which still remains open.

In order to uncover the structural features prior and during homogeneous crystal nucleation process in large scale molecular dynamics simulations without \textit{a priori}, an unsupervised learning approach \cite{Ceriotti2019} was proposed very recently \cite{Becker2022,Becker2022-2}. The method is based on topological data analysis (TDA) through persistent homology (PH) \cite{Motta2018,Carriere2015}, which emerged recently in the field of materials science as a mean to identify local atomic structure as a post-treatment in atomic scale simulations \cite{Sasaki2018,Hong2019,Hirata2020}, or experimentally in Scanning Electron Microscopy images of microstructure in aluminium alloys \cite{Adams2017,Kassab2022}. The originality of our method was to use PH as a translational and rotational invariant descriptor to encode the local structures. A  Gaussian Mixture Model (GMM) clustering method \cite{tibshirani} is then applied,  and estimated through an Expectation Maximization (EM) algorithm \cite{Dempster1977}.  This method called hereafter TDA-GMM, was shown to be successful to identify and describe the structural and morphological properties of the nuclei for various monatomic metals, namely  Al, Zr, Ta and Mg, revealing their specific nucleation pathways \cite{Becker2022,Becker2022-2}. The application of the PH to multi-component alloys as a post-treatment was the subject of very recent works on ice \cite{Hong2019} imposing specific constraints on O-H and O-O bonds, or on Pd-Si alloy \cite{Hirata2020} by removing the central atom of a given type to discriminate local structures around Pd or Si. However, in the latter case, while the topological ordering was successfully captured, the chemical short-range order (CSRO) seems to be more difficult to describe that way.

In the present work, large-scale molecular dynamics simulations (MD) are carried out  to  investigate the homogeneous nucleation and nucleation pathways of deep undercooling of Al-Ni alloys for two specific compositions, namely Al\textsubscript{50}Ni\textsubscript{50}  and Al\textsubscript{25}Ni\textsubscript{75}, that possess different underlying stable crystalline phase, namely B2 and L1\textsubscript{2}, respectively. Interestingly, Al-Ni alloys are known to be poor glass forming alloys \cite{Jakse2015,Tang2018} with small crystallization times that are reachable by brute force MD. From a methodological point of view, the PH descriptor in our TDA-GMM scheme is extended here with the objective of describing both the chemical ordering, focusing on the central atom and its distance to the first neighborhood shell, and the topological local ordering using edge-weighted persistent homology (EWPH), in a similar way as it was used in biomolecular data analysis \cite{Cang2017, Cang2018, Meng2020, Anand2020} to capture information between different elements and/or molecules.
Then a GMM is built specifically for each alloy by including in the training set samples of all possible crystalline structures \cite{MatProject,Aflow},  liquid configurations in the stable and undercooled states, and out-of-equilibrium configurations at various stages of the homogeneous nucleation process. Our results show that the nucleation is initiated first from the chemical ordering, followed by progressive bond-orientational ordering of the underlying crystal phase. Our findings further indicate that the nucleation pathway depends on composition, with Al\textsubscript{50}Ni\textsubscript{50} displaying a single step nucleation with the emergence of B2 short-range order, and for Al\textsubscript{25}Ni\textsubscript{75} a step-wise nucleation toward the L1\textsubscript{2} phase with the emergence of fcc-, hcp-, and bcc-type polymorphs in a first stage.

\section{Unsupervised learning approach for alloys}
\label{Sec:TechnicalBackground}

\subsection{Molecular dynamics simulations}
\label{Section:DataProduction}

\begin{figure}[t]
	\centering
	\includegraphics[width=1.0\textwidth]{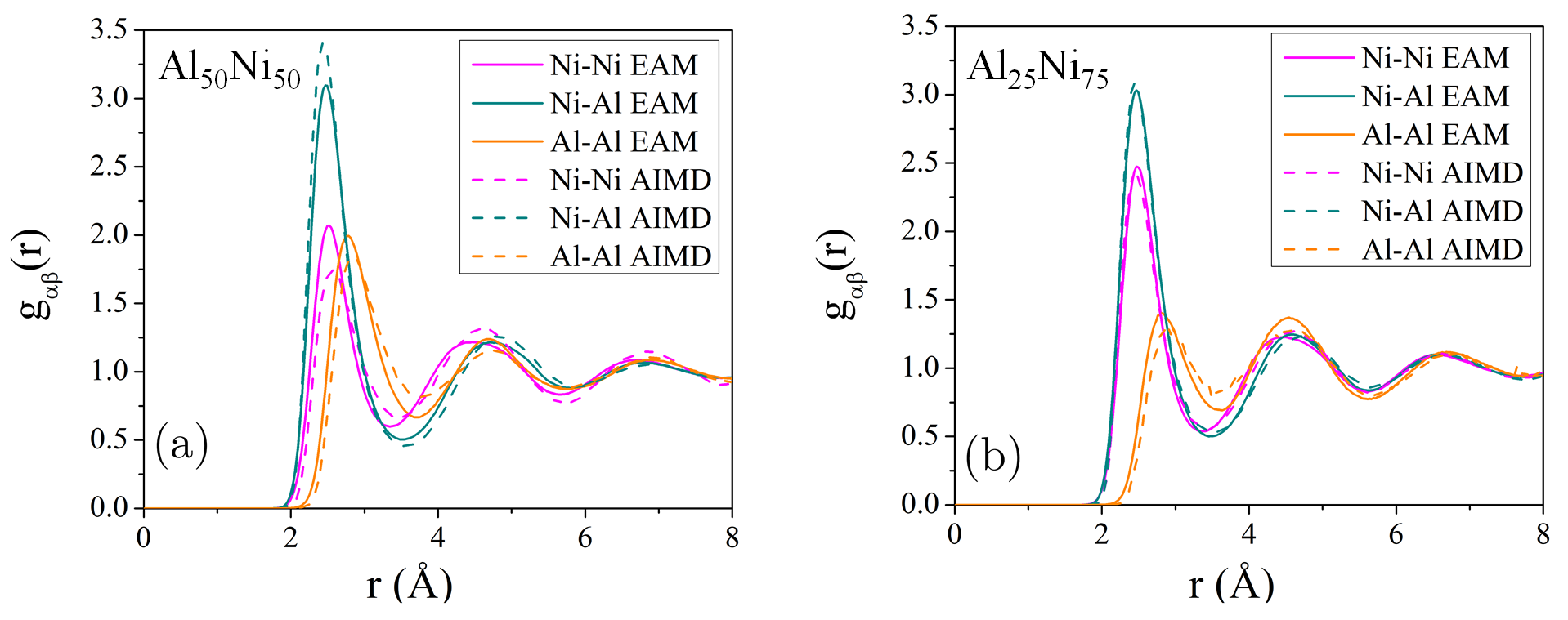}
	\caption{Partial pair-correlation functions for Al\textsubscript{50}Ni\textsubscript{50} (a) and Al\textsubscript{25}Ni\textsubscript{75} (b) at $T=1750$~K. Classical MD simulation with the EAM potential of Ref. \cite{Mishin2009} (solid lines) are compared with AIMD simulations of Ref. \cite{Pasturel2015,Pasturel2015-2} (dashed lines). }
 \label{fig:gpartiels}
\end{figure}

In order to track homogeneous nucleation in Al-Ni alloys, large-scale MD simulations were performed using the \textsc{lammps} code \cite{LAMMPS}. A number of atoms  $N=$ 1 024 000 atoms  were considered, of which 512 000 of each specy for Al\textsubscript{50}Ni\textsubscript{50}, and 256 000 Al and 768 000 Ni for Al\textsubscript{25}Ni\textsubscript{75}. They were placed randomly in a cubic simulation box subject to the standard periodic boundary conditions (PBC) in the three directions of space. Interactions were taken into account through the semi-empirical potentials of Purja and Mishin \cite{Mishin2009} based on the embedded atom model. Equations of motion are solved numerically with Verlet's algorithm in its velocity form with a time step of $1$ fs \cite{Frenkel1996,Allen2017}. Control of the thermodynamic parameters was done with the Nos\'e-Hoover thermostat and barostat \cite{Nose1984,Martyna1994} and all the simulations were conduced at constant pressure. Nucleation events were produced in undercooled states at constant temperature. Properties of each alloy are summarized in Table~\ref{Tab:TTT}, and Fig. \ref{fig:gpartiels} shows a comparison between the classical MD simulation with the EAM potential \cite{Mishin2009} and our previous \textit{ab initio} MD simulations \cite{Pasturel2015,Pasturel2015-2} for the two compositions considered here at $T=1795$ K taking the same densities. A reasonably good agreement can be seen for both alloys, and more importantly, an excellent match of the first peaks position for the three partials is found, indicating that the Al-Al, Al-Ni and Ni-Ni bond lengths are well reproduced. Especially, the first peak of Al-Ni partials are well reproduced indicating that the strong chemical affinity between the two species is well predicted and that the EAM potential is reliable. 

\begin{table}[t]
	\centering
	\begin{tabular}{lccccccccc}
		\hline\hline
		Alloy & $N$\textsubscript{Al} &  $N$\textsubscript{Ni} & {$T_{L}$ (K)} & {$T_{g}$ (K)} &{$T_{\textrm{iso}}$ (K)} & {$T_{rg}$} &{$\Delta{T}$} & {$Q$ (K.s\textsuperscript{-1})} & {$Q_{c}$ (K.s\textsuperscript{-1})} \\
		\hline
		Al\textsubscript{50}Ni\textsubscript{50} & $5.12\times10^5$ & $5.12\times10^5$ & 1780 & 915 & 1150 & 0.51 & 0.35 & $5.0\times 10^{12}$ & $1.4\times 10^{12}$ \\
		Al\textsubscript{25}Ni\textsubscript{75} & $2.5\times10^5$ & $7.5\times10^5$ & 1678 & 815 & 1050 & 0.49 & 0.37 & $1.0\times 10^{11}$ & $9.0\times 10^{10}$ \\
		\hline\hline
	\end{tabular}
	\caption{Characteristic parameters of MD simulations. $N$\textsubscript{Al} and  $N$\textsubscript{Ni} are the number of Al and Ni in the simulation boxes. Liquidus temperatures $T_{L}$ are taken from Ref. \cite{Mishin2009}. $T_g$ represents the glass transition temperature, $T_{\textrm{iso}}$ the chosen isotherm for the analysis of the nucleation events, $T_{rg}=T_{g}/T_{m}$, $\Delta T =(T_{m}-T_{iso})/T_{m}$, $Q$ the cooling rate, and $Q_{c}$ the critical cooling rate, inferred from the nose of the TTT curves.}
	\label{Tab:TTT}
\end{table}

\begin{figure}[t]
	\centering
	\includegraphics[width=0.9\textwidth]{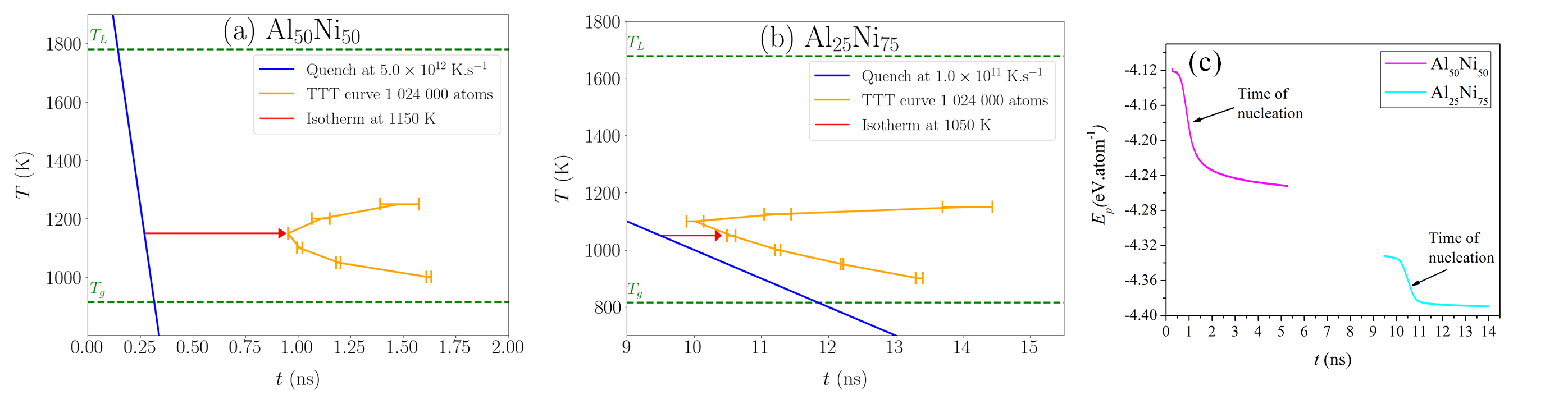}
	\caption{TTT curves in a temperature range near the nose for Al\textsubscript{50}Ni\textsubscript{50} (a) and Al\textsubscript{25}Ni\textsubscript{75} (b). (c) Potential energy along $T_{\textrm{iso}}$ along a nucleation process as a function of time for both alloys. The sharp drop of the potential energy marks the nucleation events and the nucleation time is taken as the time for which 30 \% of the atoms are in a local crystalline order according to the CNA analysis.}
 \label{fig:1}
\end{figure}

For the study of the crystal nucleation, the alloys were equilibrated in the liquid state, at $T=2500$ K and $T=2000$ K respectively for Al\textsubscript{50}Ni\textsubscript{50} and Al\textsubscript{25}Ni\textsubscript{75}, far above their liquidus temperatures $T_{L}$ (see Table \ref{Tab:TTT}). They were subsequently brought in the undercooled states, using a ramp of temperature with a cooling rate $Q=5.0\times10^{12}$ K.s\textsuperscript{-1} and $Q=1.0\times10^{11}$ K.s\textsuperscript{-1} respectively, to below the glass transition $T_{g}$ to avoid crystallization during the quench. The resulting observed values of $T_{g}$ are given in Table \ref{Tab:TTT}. Configurations are recorded during the quench by temperature steps of $50$ K. They were used to determine the time-temperature-transformation (TTT) curve in the vicinity of the nose region, as shown in Fig.~\ref{fig:1} for both alloys. For each temperature, nucleation time was determined from a sharp drop of the energy as was done in our preceding work \cite{Becker2022-2}, and is averaged over five independent runs where initial velocity distribution were randomized.

\subsection{Data preparation for the training set}
\label{Section:DataPreparation}

For the purpose of monitoring the homogeneous nucleation process with the TDA-GMM method, a number of $4$ configurations of interest along $T_{iso}$ are chosen for each alloy, spanning from the undercooled liquid state up to the poly-crystalline state after the impingement of the nuclei. Each of these configurations is first brought in its inherent structure (IS) by applying a conjugate gradient minimization in the corresponding local minimum on the potential energy surface \cite{Stillinger1982}. 

The next step it to construct a training set from these IS configurations. From the million atoms of each IS configuration, local atomic structures are subsampled to construct a training set for the learning, using the procedure developed in \cite{Becker2022-2}. A local structure is composed of all atoms around a central atom within a sphere of radius corresponding to the second minimum of $g_{\alpha\beta}(r)$ with $\alpha,\beta\in\{Al,Ni\}$. It can be seen from Fig. \ref{fig:gpartiels} that this radius, taken as $5.7$~\AA{} and $5.5$~\AA{} respectively for Al\textsubscript{50}Ni\textsubscript{50} and Al\textsubscript{25}Ni\textsubscript{75}, is similar for the three partials at the same composition.  This results in around $5600$ structures per IS configuration, taking care that the proportion of the central atom species is in agreement with the chemical composition of the corresponding alloy. Finally, to guarantee that the TDA-GMM model is able to detect existing crystalline polymorphs, local structures from all known Al-Ni crystal structures \cite{MatProject,Aflow} (whose description are given in Table \ref{Tab:CrystalsAlNi}) were added through the perfect crystal structures but also  random distortions of them modeled by a Gaussian noise (see Supplementary Information \cite{SM} for more details). Finally, the two training sets for Al\textsubscript{50}Ni\textsubscript{50} and Al\textsubscript{25}Ni\textsubscript{75} contain respectively $61 608$ and $61 775$ local structures.

\begin{table}[h!]
	\begin{tabular}{lrrr}
		\hline\hline
	    Id & \makebox[2cm][r]{Composition} & \makebox[6cm][r]{Number of sites} & \makebox[4cm][r]{Sample size}\\
		\hline
		mp-622209 & Al\textsubscript{3}Ni\textsuperscript{\textit{S}} & 2\textsuperscript{Al} - 1\textsuperscript{Ni} & 3000 \\
		mp-1057 & Al\textsubscript{3}Ni\textsubscript{2}\textsuperscript{\textit{S}} & 2\textsuperscript{Al} - 1\textsuperscript{Ni} & 3000 \\
		mp-16514 & Al\textsubscript{3}Ni\textsubscript{5}\textsuperscript{\textit{S}} & 2\textsuperscript{Al} - 3\textsuperscript{Ni} & 5000 \\
		mp-16515 & Al\textsubscript{4}Ni\textsubscript{3}\textsuperscript{\textit{S}} & 2\textsuperscript{Al} - 1\textsuperscript{Ni} & 3000 \\
		mp-1487 & AlNi\textsuperscript{\textit{S}} & 1\textsuperscript{Al} - 1\textsuperscript{Ni} & 2000 \\
		mp-2593 & AlNi\textsubscript{3}\textsuperscript{\textit{S}} & 1\textsuperscript{Al} - 1\textsuperscript{Ni} & 2000 \\
		mp-1183232 & AlNi\textsubscript{3}\textsuperscript{\textit{M}} & 1\textsuperscript{Al} - 1\textsuperscript{Ni} & 2000 \\
		mp-1025044 & AlNi\textsubscript{2}\textsuperscript{\textit{M}} & 1\textsuperscript{Al} - 2\textsuperscript{Ni} & 3000 \\
		mp-1229048 & Al\textsubscript{2}Ni\textsubscript{3}\textsuperscript{\textit{M}} & 2\textsuperscript{Al} - 3\textsuperscript{Ni} & 5000 \\
		mp-672232 & AlNi\textsubscript{3}\textsuperscript{\textit{M}} & 1\textsuperscript{Al} - 2\textsuperscript{Ni} & 3000 \\
		mp-1228713 & Al\textsubscript{4}Ni\textsubscript{15}\textsuperscript{\textit{M}} & 2\textsuperscript{Al} - 4\textsuperscript{Ni} & 6000 \\
		mp-1228854 & AlNi\textsuperscript{\textit{M}} & 1\textsuperscript{Al} - 1\textsuperscript{Ni} & 2000 \\
		\hline\hline
			\end{tabular}	
			\caption{Crystalline structure given by the Materials Project \cite{MatProject}, composition and  number of distinct sites. Superscripts (\textit{S}) and (\textit{M}) correspond respectively to a stable and metastable composition and superscripts (Al) and (Ni) to the two types of central atom. For each distinct site, a sample of $1000$ randomly distorted structures have been taken.}
			\label{Tab:CrystalsAlNi}
		\end{table}

\subsection{Topological and chemical descriptors}

In our previous works for pure elements \cite{Becker2022,Becker2022-2}, homogeneous crystal nucleation pathways were successfully characterized using an unsupervised learning approach (TDA-GMM), in which local structures were encoded by persitent diagrams (PD) via persistence homology, a main tool in TDA. A topological vector is built  from a PD $D$: each  coordinate is associated to a pair of points $(x,y)$ for a fixed level of homology, except the infinite point, and is calculated by 
\begin{equation}
\label{eq:desc}  
m_D(x,y) = \min \{\|x-y\|_\infty, d_\Delta(x), d_\Delta(y)\},
\end{equation}
where $d_\Delta(\cdot)$ denotes the $\ell^\infty$ distance to the diagonal. The coordinates are sorted by decreasing order, with a vector length equal to the number of points in the PD $D$ (for details we refer to \cite{Carriere2015,Becker2022-2}). 

For alloys, our descriptor is composed of two parts, one centered on the chemical information and the other on the topological information. 

The first part is a chemically oriented radial information encoding the chemical ordering which is a very important aspect for the short-range order of alloys \cite{Royall2015,Jakse2008,Pasturel2017}. More precisely, it is composed of two ingredients. The first one is the vector of the distances from the central atom to the atoms of the first neighbourhood shell and of the same type as the central one, sorted in a decreasing order. The second one is the vector of the signed distances from the central atom to the atoms again of the first neighbourhood shell but  of a different type from the central one, sorted again in a decreasing order. The sign of the distance from the central particle and a given neighbouring particle is chosen from a choice of a fixed total ordering on the type of atoms: if the neighboring atom is of a higher type than the central atom, the sign is positive and it is negative otherwise.

The second part of the descriptor is the concatenation of, for each type of atom, one topological vector as in our previous work with only the neighbouring atoms of the considered type. Only homology of dimension $1$ is computed, being the only one discriminating. Those choices to reduce the dimension are motivated by good performance on the reduced vector, as illustrated in Supplementary Information \cite{SM}.

Figure \ref{fig:desc} gives a synthetic overview of the descriptor in the case of a binary alloy. A more detailed description and motivation are given in Supplementary Information \cite{SM}.

\begin{figure}[t]
	\centering
	\includegraphics[width=0.9\textwidth]{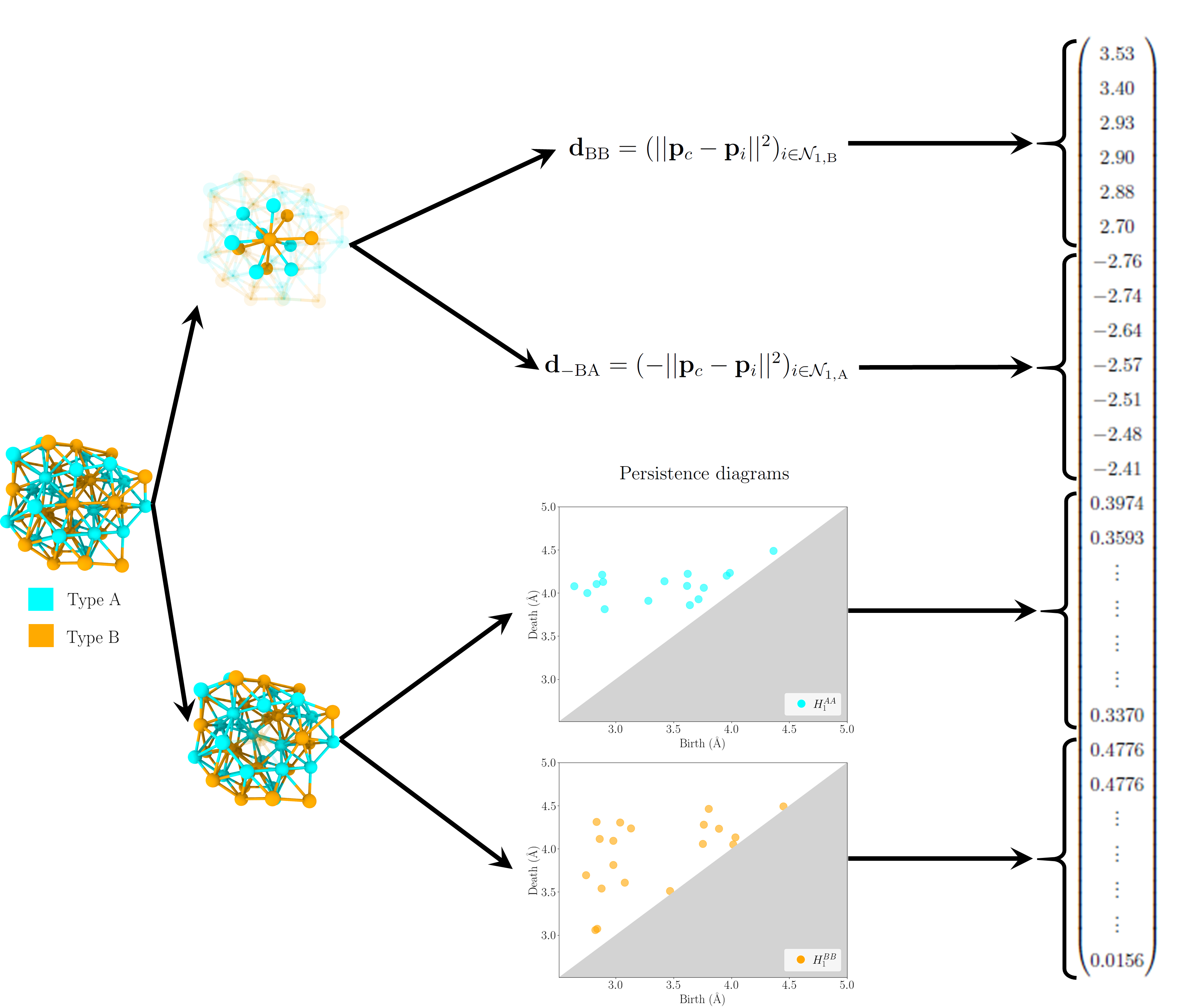}
	\caption{Schematic view of the descriptor for a binary local atomic structure with type A (cyan) and type B (orange) atoms. A slice view of the structure is provided on the left, and decomposed onto two parts using  transparency to distinguish the atoms and the bounds that are in each part of the description. For the chemically oriented vector (see text), as the type of the central atoms is B, the two parts are $\mathbf{d}_{\text{BB}}$ and $\mathbf{d}_{-\text{BA}}$ where $\mathbf{p}_{c}$ is the position of the central atom and $\mathbf{p}_{i\in\mathcal{N}_{1,\text{A}}}$ and $\mathbf{p}_{i\in\mathcal{N}_{1,\text{B}}}$ are, respectively, the position of the atoms in the first neighbors of type A or B. The two  components of the topological part, for each persistence diagram, are computed using Eq. \ref{eq:desc}.}
 \label{fig:desc}
\end{figure}

\subsection{Gaussian mixture modeling}
\label{Section:GaussianMixtureModeling}
As in the TDA-GMM method introduced in our previous work \cite{Becker2022}, a GMM is used to cluster the descriptors, with diagonal covariance matrices. 
An EM algorithm \cite{Dempster1977} is used to estimate the parameters, through the \texttt{scikit-learn} package \cite{Pedregosa2011}. The model is initialized with the best model among $3000$ $k$-means runs.  An elbow criterion on the likelihood is used to select the number of clusters, which leads to exactly $50$ components for the two datasets. 

As a posterior validation of the GMM model, more than $99.5$\%  of the  structures not used in the training set are assigned to a cluster with a posterior probability larger than $0.999$. This gives a clear distinction between clusters, and the remaining observations,  coming     from the simulated configurations of nucleation, still have posterior probabilities higher than $0.5$. 

\begin{figure}[h!]
	\centering
	\includegraphics[width=1\textwidth]{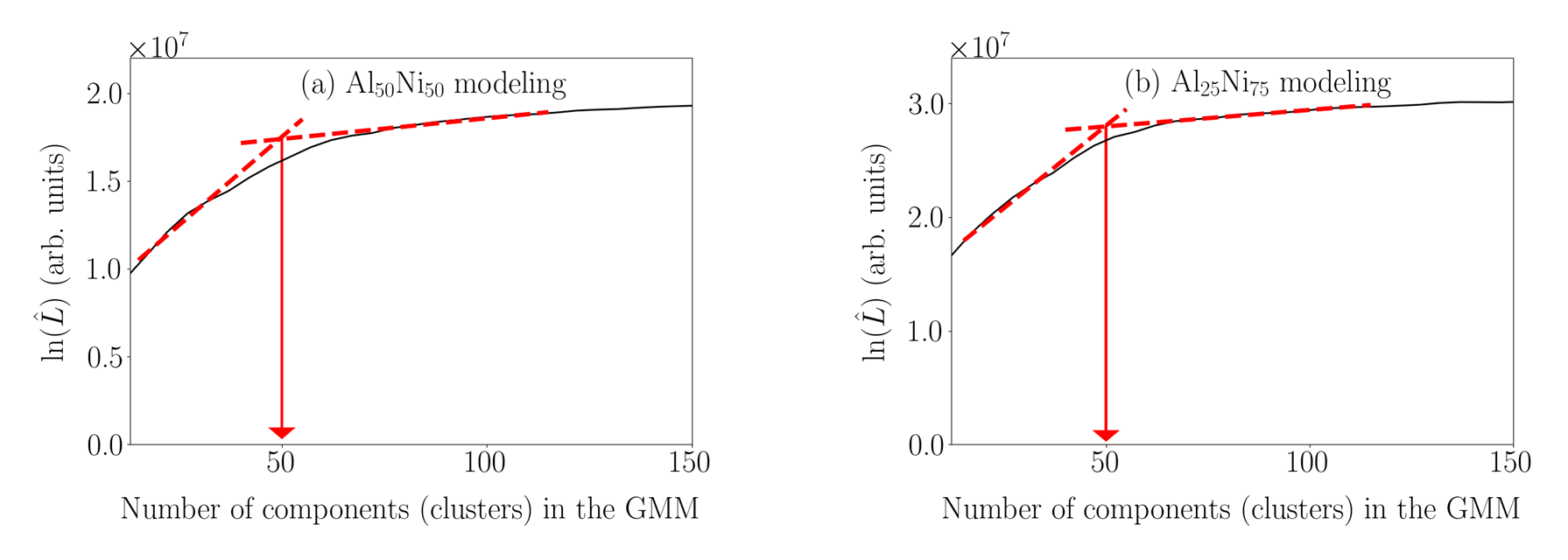}
	\caption{\label{fig:3}Log-likehood functions for the Al\textsubscript{50}Ni\textsubscript{50} (a) and Al\textsubscript{25}Ni\textsubscript{75} (b) modeling, with the tangents to the curve in the vicinity of the inflection point symbolized by the red dotted lines.}
\end{figure}

\section{Results and discussion}
\label{Sec:ResultsAndDiscussion}

\begin{figure}[t]
		\centering
		\includegraphics[width=1.0\textwidth]{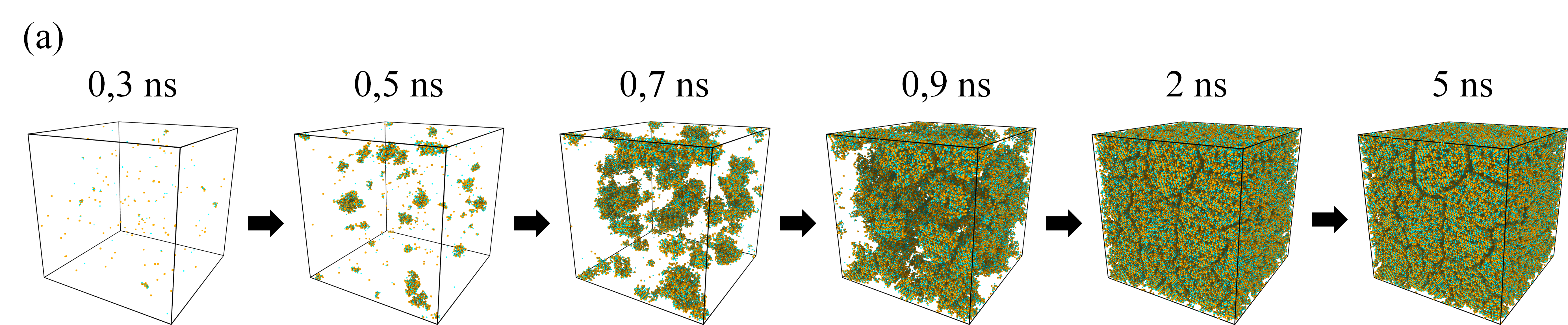}
		\includegraphics[width=1.0\textwidth]{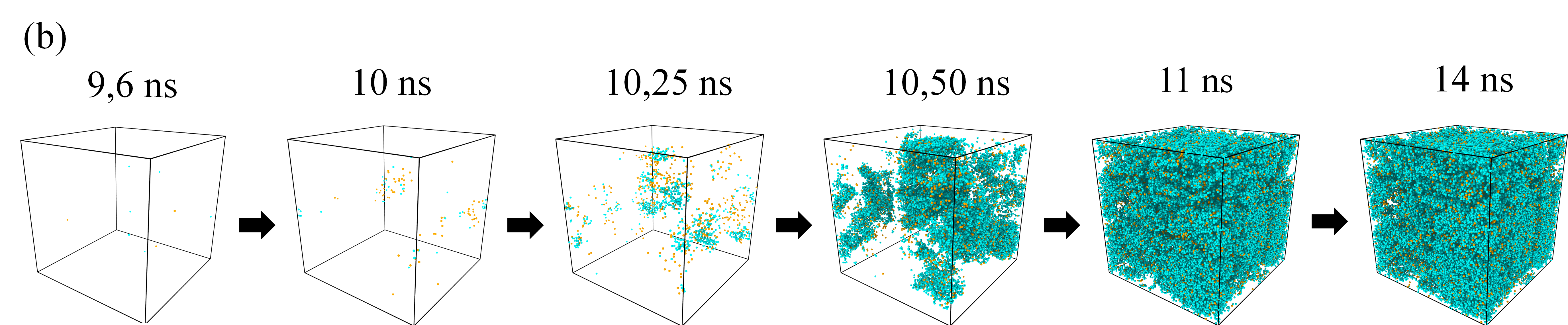}
		\caption{\label{fig:NucleationAlNi}Snapshots of MD simulations at various stages of the homogeneous nucleation for (a) Al\textsubscript{50}Ni\textsubscript{50} and (b) Al\textsubscript{25}Ni\textsubscript{75}. Only Al (orange) and Ni (cyan) atoms belonging to clusters with B2 and L1\textsubscript{2}, respectively, are drawn (see text).}
	\end{figure}
 
    The homogeneous nucleation pathways in Al\textsubscript{50}Ni\textsubscript{50} and Al\textsubscript{25}Ni\textsubscript{75} explored by MD are illustrated in Fig. \ref{fig:NucleationAlNi} by drawing simulation snapshots at specific times during the process in deep undercooling conditions close to the nose of the TTT at $T=1150$~K and $T=1050$~K, respectively. 
    For each composition, all atoms of the simulation box are first associated to one of the $50$ clusters of the trained TDA-GMM model, as described in the previous Section \ref{Section:GaussianMixtureModeling}. 
    The evolution of the proportion of local structures in each cluster is monitored during the whole process, and those for which a significant increase is detected are deemed to participate to the nucleation, without prior knowledge of their orientational and chemical orderings (see also Supplementary Information \cite{SM}). 
    For Al\textsubscript{50}Ni\textsubscript{50}, $16$ clusters with increasing proportions are seen, and they all display a bcc ordering as analyzed by the CNA \cite{Honeycutt1987,Faken1994} with various chemical orderings in terms of the Cowley parameter \cite{Cowley1950}.
    For Al\textsubscript{25}Ni\textsubscript{75} the situation appears to be more complex with $22$ clusters having an increasing proportion of local structures with fcc, hcp, and bcc structures with various chemical ordering. 
    The relevant clusters proportions in both cases are gathered in Table SIII and SIV in Supplementary Information \cite{SM}.     
    In Fig. \ref{fig:NucleationAlNi}, atoms are drawn only if they belong to clusters having a B2 structure for Al\textsubscript{50}Ni\textsubscript{50} with Cowley parameter $\alpha_{1}=-0.1429$, and L1\textsubscript{2} structure for Al\textsubscript{25}Ni\textsubscript{75} with $\alpha_{1}=-0.333$, which correspond to their respective underlying crystalline phase. 
    This highlights clearly for both alloys the homogeneous nucleation stage up to the impingement followed by the grain growth stage, also called the solidification. the latter is a much slower process that can not be completely captured by brute force MD. 
    For Al\textsubscript{50}Ni\textsubscript{50} nucleation starts around $0.3$~ns with multiple nucleation events up to $1$~ns where solidification starts. 
    The same holds for Al\textsubscript{25}Ni\textsubscript{75} $\alpha_{1}=-0.333$ with nucleation appearing at $10$~ns and a solidification stage starting $1.75$~ns later. 

\begin{figure}[t]
	\centering
	\includegraphics[width=0.49\textwidth]{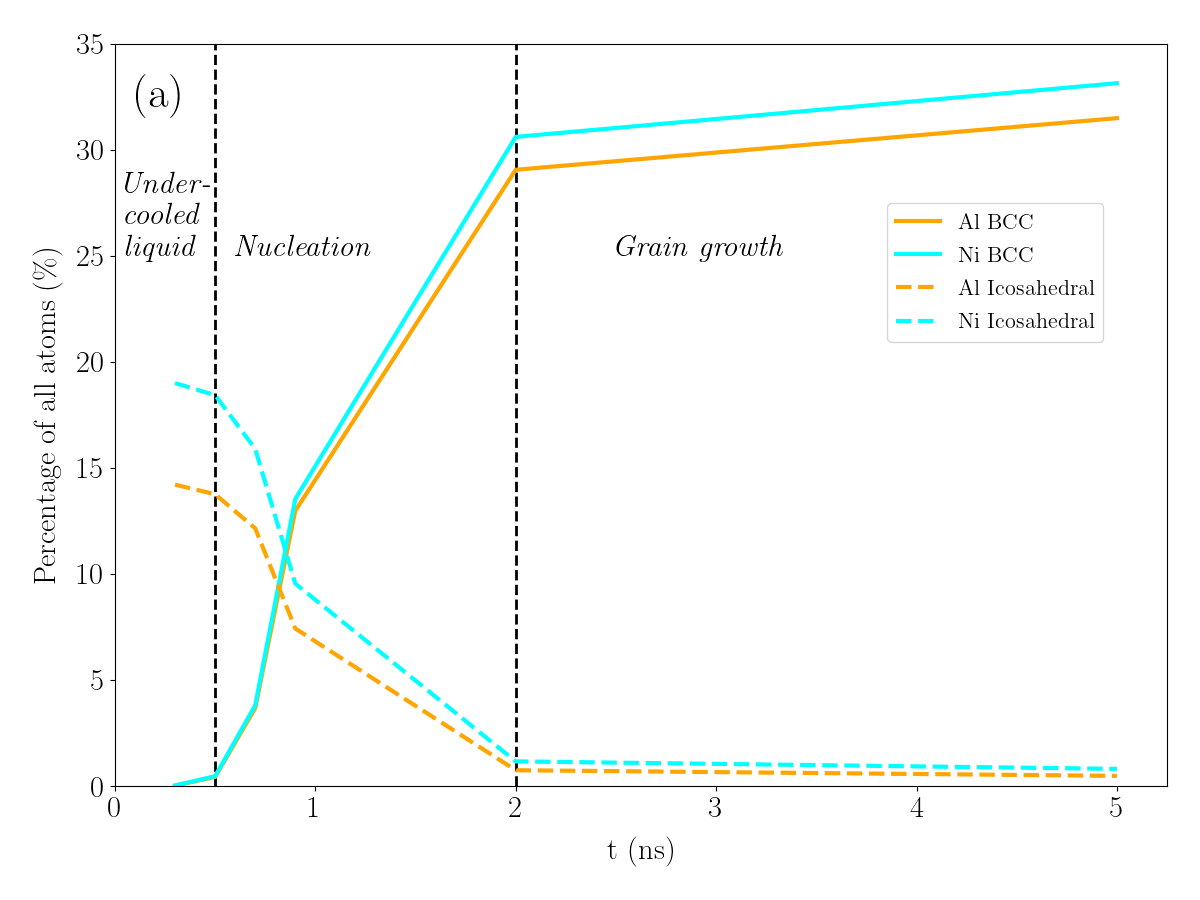}
	\includegraphics[width=0.49\textwidth]{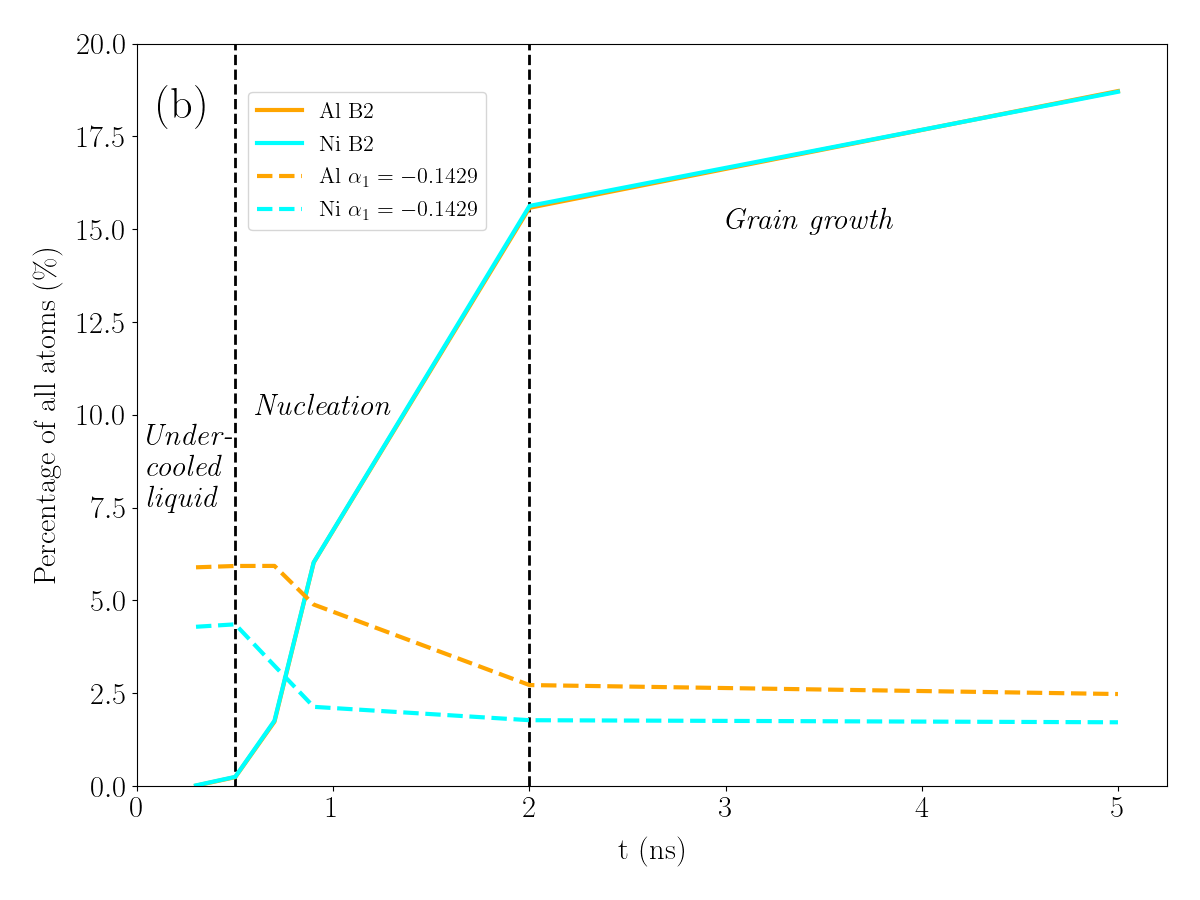}
	\caption{\label{fig:OrderingsAl50Ni50} Evolution of the crystalline and five-fold symmetry of Al\textsubscript{50}Ni\textsubscript{50} during nucleation at $T=1150$ K(a) for the bond-orientational ordering, and (b) chemical short-range order with  $\alpha_{1} = -0.1429$ corresponding to a B2 signature.}
\end{figure}

The bond-orientational and chemical orderings during nucleation is first examined for Al\textsubscript{50}Ni\textsubscript{50}. 
In Fig. \ref{fig:OrderingsAl50Ni50}(a) the cumulative proportions of local structures in clusters having a bcc structure are shown, as well as those having a dominant five-fold symmetry (FFS) local ordering, also called icosahedral ordering, distinguishing Al and Ni centered ones. 
It is worth mentioning that local structures with fcc or hcp orderings are negligibly small for this composition so they are not displayed in the plot.
In the undercooled liquid, the FFS ordering is slightly higher around Ni atoms than Al ones, which is consistent with our former AIMD simulations \cite{Pasturel2015-2}. 
The onset of nucleation is characterized by a strong increase of bcc ordering, which slows down in the grain growth stage, and a simultaneous strong decrease of the FFS short-range order. 
While the bcc order is preponderant around Ni, Fig.  \ref{fig:OrderingsAl50Ni50}(b) shows that both species are equally involved in the B2 ordering. 
More interestingly, Fig.  \ref{fig:OrderingsAl50Ni50}(b) indicates that in the liquid prior to nucleation, a significant chemical ordering, corresponding to the B2 with $\alpha_{1} = -0.1429$, exists in clusters with disordered local structures 
This is in favor of a scenario in which the chemical ordering of the phase is already present before the bond-orientational ordering sets in.
As a matter of fact, Table SIII in Supplementary Information \cite{SM} shows that clusters with B2 local structures at the early stages of nucleation have a higher proportion.

\begin{figure}[t]
	\centering
	\includegraphics[width=0.49\textwidth]{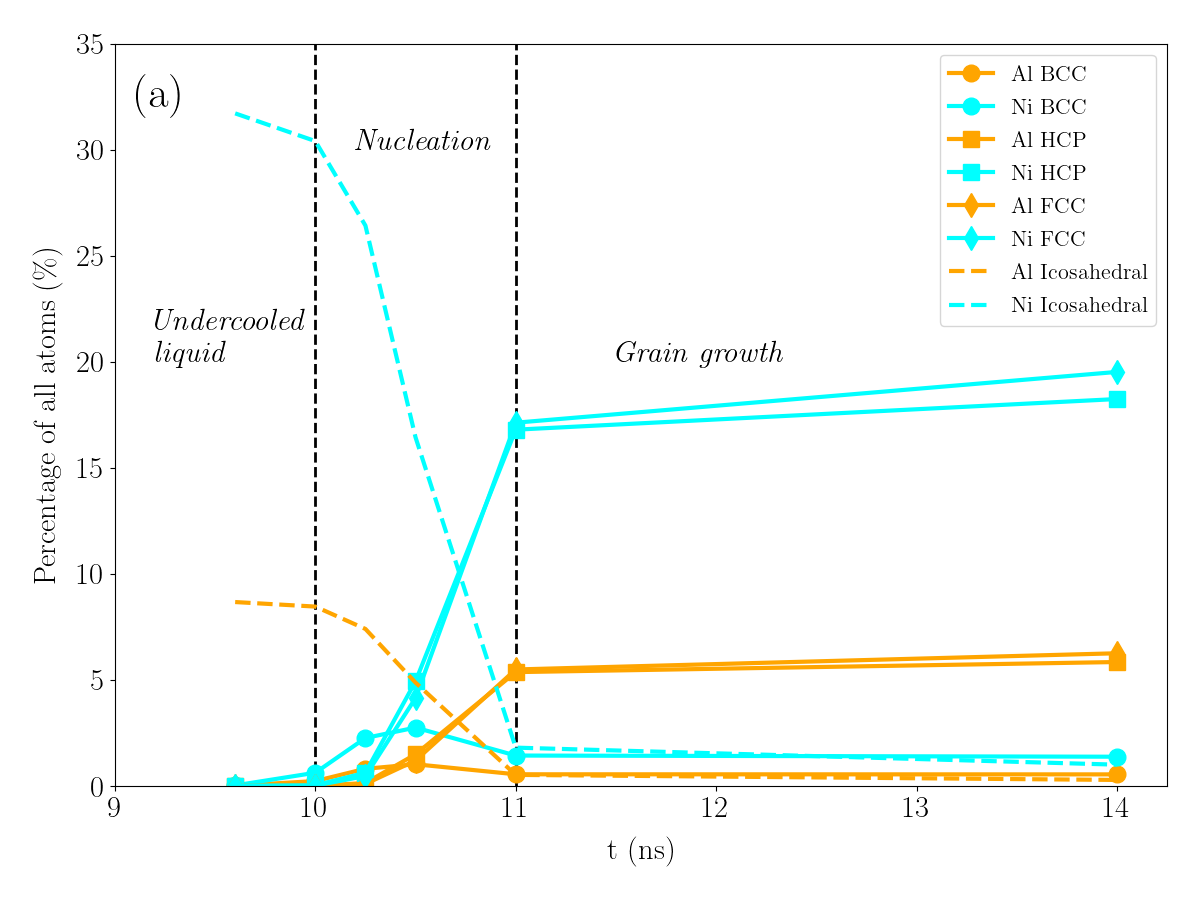}
	\includegraphics[width=0.49\textwidth]{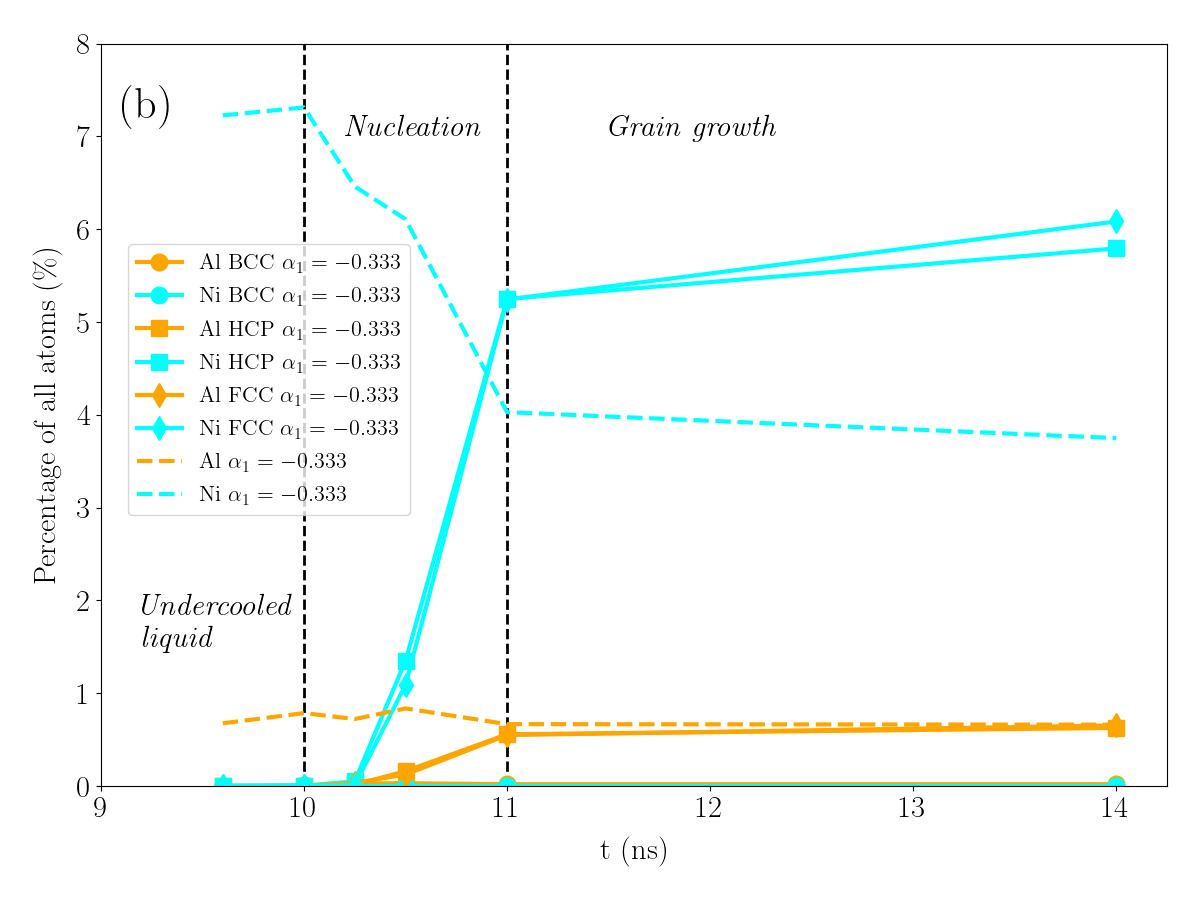}
	\caption{\label{fig:OrderingsAl25Ni75} Evolution of the crystalline and five-fold symmetry of Al\textsubscript{25}Ni\textsubscript{75} during nucleation at $T=1050$ K (a) for the bond-orientational ordering, and (b) chemical ordering with  $\alpha_{1} = -0.333$ corresponding to a L1\textsubscript{2} signature.}
\end{figure}

For Al\textsubscript{25}Ni\textsubscript{75}, the bond-orientational and chemical orderings are displayed in Fig. \ref{fig:OrderingsAl25Ni75} as for the equiatomic composition above. 
Proportions of local structures having a crystalline order are composed of various polymorphs, namely bcc, hcp and fcc as can be seen in Fig. \ref{fig:OrderingsAl25Ni75}(a). 
It should be noticed that the relative proportions of Al and Ni in all the bond-orientational orderings are consistent with the alloy composition. 
The FFS is more pronounced than for Al\textsubscript{50}Ni\textsubscript{50}, and dominant around Ni in the undercooled liquid, which is again consistent with AIMD results \cite{Pasturel2015-2}, and essentially vanishes during the nucleation stage. 
The onset of nucleation is characterized by a two-stage process in which the bcc ordering is selected first. 
It  transforms back to the fcc and hcp orderings in a second stage when the fcc and hcp experience a strong increase  in equal proportions. 
They finally become more steady after impingement when the grain growth stage sets in. 
A closer inspection shows that the hcp local order participates in numerous stacking fault in the nuclei as well as in grain boundaries as shown in Fig. S4 in Supplementary Information \cite{SM}. 
A significant chemical ordering corresponding to the L1\textsubscript{2} with $\alpha_{1} = -0.333$ Fig.  \ref{fig:OrderingsAl25Ni75}(b) preexists in the undercooled liquid, which is again an indication for this composition, that the chemical ordering of the underlying phase precedes the bond-orientational ordering at the onset of nucleation (see also Table SIV in Supplementary Information \cite{SM}).

\begin{figure}[t]
	\centering
	\includegraphics[width=1\textwidth]{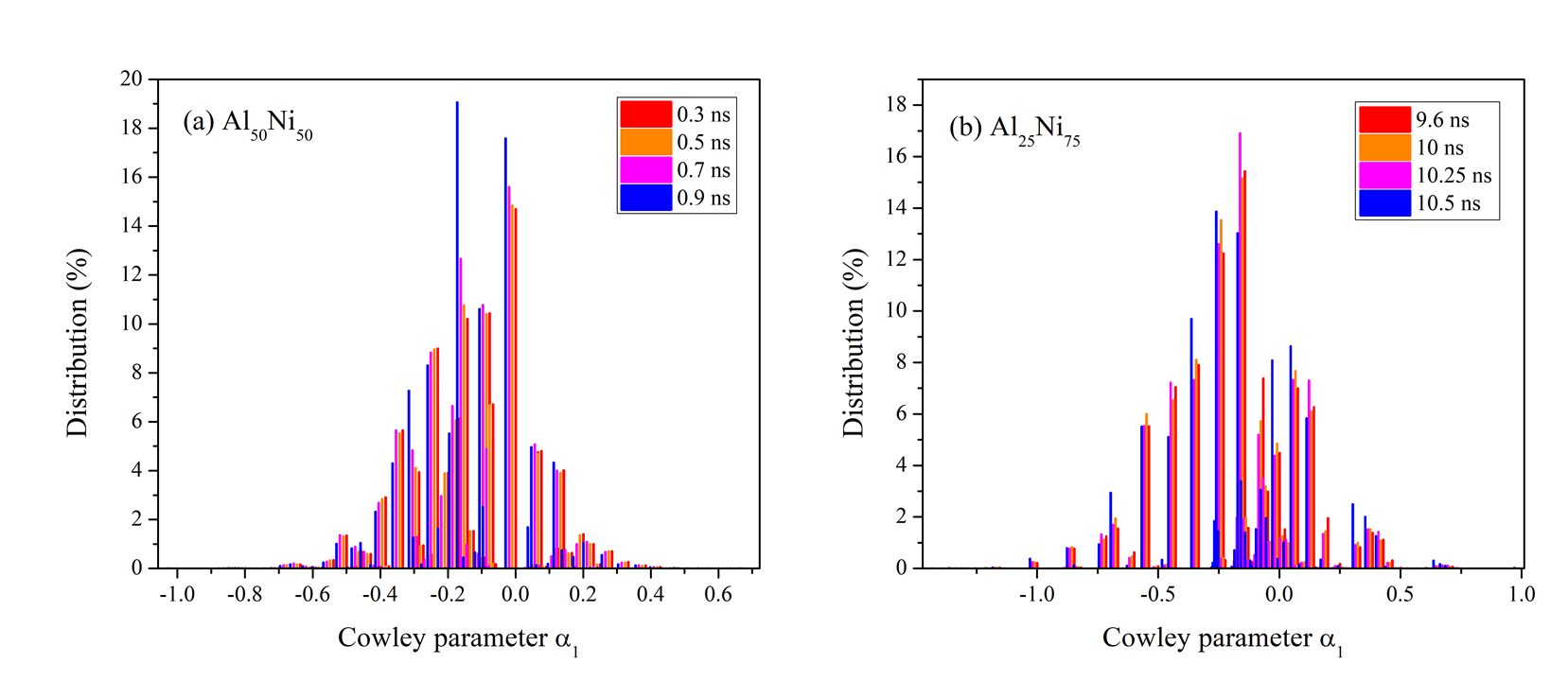}
	\caption{\label{fig:CSROAl25Ni75} Distribution of the chemical short-range order around all atoms for the configurations of  Fig. \ref{fig:NucleationAlNi} during the nucleation phase for (a) Al\textsubscript{50}Ni\textsubscript{50} and (b) Al\textsubscript{25}Ni\textsubscript{75}. The values $\alpha_{1} = -0.333$ $\alpha_{1} = -0.1429$, and $\alpha_{1} = 0$ correspond to  L1\textsubscript{2}, B2, and disordered  signatures, respectively.}
\end{figure}

Finally, our unsupervised learning approach enables us to show that the two compositions studied possess two distinct nucleation pathways, namely a single step nucleation for Al\textsubscript{50}Ni\textsubscript{50}, and a two-step nucleation for Al\textsubscript{25}Ni\textsubscript{75} following Ostwald's rule \cite{Ostwald1897}. 
In both cases, the preexisting chemical short-range order of the underlying phase precedes the bond-orientational ordering. 
This can be seen from the local chemical ordering distributions in Fig.\ref{fig:CSROAl25Ni75}. For Al\textsubscript{50}Ni\textsubscript{50}, beside the disordered peak, a preponderant peak is seen around the B2 value of $\alpha_{1}$ before and during the whole nucleation stage. For  Al\textsubscript{25}Ni\textsubscript{75}, the distributions show a peak around the B2 at the early stage that subsequently decrease at the expense of a peak at the L1\textsubscript{2} value of $\alpha_{1}$.
For Al\textsubscript{50}Ni\textsubscript{50}, our results are consistent with previous MD simulations with the same EAM potential \cite{Orihara2020}, in which a single step towards the B2 phase was observed albeit with a lower bcc ordering. 
It was proposed \cite{Orihara2020} that the two-step nucleation observed in pure Ni starting with the bcc local ordering could be at the origin  of the B2 polymorph selection in this alloy. 
Our results are not in favor of this scenario as they rather indicate a symmetric role of Al and Ni that can be explained by their strong affinity and similar diffusion coefficients at that composition \cite{Pasturel2015-2}.
Moreover, it was shown in a preceding work \cite{Becker2022} that homogeneous nucleation in pure Al is characterized by a single step with an initial fcc ordering. This is an additional counterargument. 
The situation is different for Al\textsubscript{25}Ni\textsubscript{75} for which the two-step nucleation can come from that of pure Ni \cite{Orihara2020}. 
Then, our results might indicate for this composition  that the two-step pathway is a combined effect of that of pure Ni and of the preexisting chemical ordering of the underlying phase at the onset of nucleation.  
 
\section{Conclusion}
In conclusion, this study was devoted to the investigation of homogeneous nucleation and nucleation pathways of deep undercooled Al\textsubscript{50}Ni\textsubscript{50}  and Al\textsubscript{25}Ni\textsubscript{75} alloys by means of large-scale MD simulations. 
For this purpose an unsupervised topological and chemical learning approach was successfully extended to alloys from our previous work based on the persistent homology developed for pure elements \cite{Becker2022,Becker2022-2}, by including the chemical ordering in the PH descriptor for building a Gaussian mixture model. 
This allowed us to show that the chemical short-range order of the underlying stable crystalline phase of both alloys preexist in the undercooled liquid prior to the onset of crystal nucleation, followed by an increase of the bond-orientational ordering. Our findings further indicate that the nucleation pathway depends on composition, with Al\textsubscript{50}Ni\textsubscript{50} displaying a single step nucleation with the emergence of B2 short-range order, and for Al\textsubscript{25}Ni\textsubscript{75} a step-wise nucleation toward the L1\textsubscript{2}  phase with the emergence of bcc-type polymorph in the first stage that might come from the combine effect of the chemical ordering and the two-stage nucleation pathway of pure Ni. 

\section*{Acknowledgement}

We acknowledge the CINES and IDRIS under Project No. INP2227/72914, as well as CIMENT/GRICAD for computational resources. This work was performed within the framework of the Centre of Excellence of Multifunctional Architectured Materials CEMAM-ANR-10-LABX-44-01 funded by the "Investments for the Future" Program. This work has been partially supported by MIAI@Grenoble Alpes (ANR-19-P3IA-0003). Fruitful discussions within the French collaborative network in high-temperature thermodynamics GDR CNRS3584 (TherMatHT) and in artificial intelligence
in materials science GDR CNRS 2123 (IAMAT) are also acknowledged.

\renewcommand{\thefigure}{S\arabic{figure}}
\setcounter{figure}{0}
\renewcommand{\thetable}{S\Roman{table}}
\setcounter{table}{0}
\renewcommand{\theequation}{S\arabic{equation}}
\setcounter{equation}{0}

\newpage
\begin{center}
	{\Large Supplemental Information}
\end{center}

\section*{Data preparation: construction of a set of local crystalline structures}

We built a set of local crystalline structures up to the second shell of neighbors from the available known unit cells of AlNi crystal alloys \cite{MatProject}. Each unit cell was reproduced in the three directions of space to obtain virtual boxes of about 30 000 atoms. A Gaussian noise, centered and with a standard deviation of $0.01$ \AA~was then applied to the atomic positions to perturb the crystal lattices without losing the structural signature of the different sites. We have checked that the partial radial distribution functions and the common neighbor analysis (CNA) signatures \cite{Honeycutt1987} remained the same between the noisy and initial crystalline configurations.
The CNA was performed here with a fixed cut off radius $r=3.5$ \AA~\footnote{In the case of the CNA performed for the structures in the MD configurations studied in the main text, we took three cut-off radii, one different for each of the possible chemical bonds: $r_{\text{Al}-\text{Al}}=3.57$ \AA, $r_{\text{Al}-\text{Ni}}=3.49$ \AA, $r_{\text{Ni}-\text{Ni}}=3.33$ \AA~for Al\textsubscript{50}Ni\textsubscript{50}; and $r_{\text{Al}-\text{Al}}=3.38$ \AA, $r_{\text{Al}-\text{Ni}}=3.35$ \AA, $r_{\text{Ni}-\text{Ni}}=3.24$ \AA~for Al\textsubscript{25}Ni\textsubscript{75}.}, using the Python implementation in the software \textsc{ovito} \cite{Stukowski2010}.
We identified for each composition the distinct sites with two shells of neighbors  by discriminating autonomously the atomic structures with a CNA signature using the atomic bonds of the central atom and the first neighbors of the same type as the latter,  normalized with  the total number of bounds (central atom and first shell of neighbors). We used the Python package \texttt{pyscal} \cite{Menon2019} on each full configuration to  extract the  coordinates of the atoms. Finally, a sample of  1000 randomly distorted structured for each site is picked in each composition, which leads to a set composed of 39 000 structures (21 000 with a central atom Ni and 18 000 with a central atom Al).

\section*{Edge-weighted persistent homology, chemical and topological description}

In our previous work \cite{Becker2022,Becker2022-2} the idea was to encode the topological information using persistent homology. Since we are now working with an alloy, we have to take into account the different types of atoms. Inspired by \cite{Cang2017, Cang2018, Meng2020, Anand2020}, we used Edge-Weighted Persistent Homology (EWPH). The idea is to adapt the appearance of 1-simplices in the Vietoris-Rips complexes along the filtration according to the types of the extremities. 
We computed several persistent diagrams, one for each possible (non ordered) pair of types (in the case of an alloy AB, we have three types A-A, A-B, B-B) letting only appearing 1-simplices corresponding to this pair. 
More precisely, as the Vietoris-Rips complexes can be computed from the distance matrix $M={\{M_{ij}=d(x_{i},x_{j}), i,j \in \{1,\dots,\mu\}\}}$ of the point cloud $X=(x_{i})_{i=1,\dots,\mu}$, we define the components of a modified distance matrix for each non ordered pair of type $(\alpha,\beta)$ as
\begin{equation}
	\label{eqdistancematrix}
	M^{\alpha,\beta}_{ij}=
	\begin{cases}
		d(x_{i},x_{j})&\text{if $(\text{type }x_i,\text{type }x_j)=(\alpha,\beta)$ ;}\\
		\infty&\text{otherwhise.}
	\end{cases}
\end{equation}
In practice, $\infty$ corresponds to a value larger than the maximum filtration size. Thus, for a structure with two types of atoms $A$ and $B$, three modified distance matrices  are defined, which lead to three independent filtration processes, each capturing the topological information given by one of the possible chemical interactions. The generalization to higher order alloys is straightforward and the case of a single-element atomic structure corresponds to a single modified matrix, which strictly corresponds to the distance matrix of the unweighted PH. 

We can use this artefact to obtain only information from one or a set of specific atoms in the structure. In other words, it is possible to divide a structure into several substructures. 

First, to focus on the chemical information, we consider a substructure with the central atom and its   neighbors in the first neighbors shell.
We set $M_{i,j}=\infty$ in \ref{eqdistancematrix} if $x_i$ or $x_j$ is not the central atom, and if the other is not in the first neighbors shell. However, we only get information in dimension 0, and a radial information contains much more information with less parameters, which we encode in our chemically ordered radial information vector. Its first components correspond to the distances, arranged in increasing order, from the central atom to the neighboring atoms of the same type as the central one. Remaining components are the signed distances, still arranged in decreasing order, from the central atom to the atoms with a different type from the central one. In these last components, the sign is obtained by ordering the types of atoms: the distance is positive if the type of the neighboring atom is higher than the one of the central one and negative otherwise.   

We can also obtain another substructure corresponding to the second shell of neighbors by taking into account all the distances in \eqref{eqdistancematrix} except the ones between the central atom and its first neighbors,   setting $i,j\ne1$.
For each pair $(\alpha,\beta)$ of possible types (so $(A,A)$, $(A,B)$ and $(B,B)$ for a binary alloy with atoms of type $A$ and $B$) we obtain a persistent diagram. In the diagram associated to the pair $(\alpha,\alpha)$ we have points in homological dimension 0, 1 and 2. In the diagram associated to the pair $(\alpha,\beta)$ with $\alpha\neq \beta$ we have points in homological dimension 0 and 1. Indeed, there is no point in homological dimension 2 because there are only 0-simplicies and 1-simplices arising in the filtration process. Moreover,  the points in homological dimension 1 are at infinity. In the cases of NiAl alloy considered in this paper, they were several hundreds of points at infinity for a given local structure which give rise in the end to a topological vector with more than one thousand of components. This is too much to deal with our procedure and we decided to forget this part of the topological vector in our study.

The final descriptor constructed here has for first component the chemical information, and the second component corresponds to 3 PDs with $H_0$, and potentially $H_1$ and $H_2$, components.

\section*{Supervised learning to select sufficient topological information}

As described in the previous section, the full descriptor has a lot of coefficients (about 300). To have a good clustering with a GMM, we would need a huge number of local structures in the training set. As it was not the case in our dataset,  we needed to reduce the dimension of the descriptor. Also, since we are doing several PDs on the same local structures, though with different weights, it is reasonable to think that there is redundancy in the topological information we obtain. 

Using the set of local crystalline structures we constructed, we wanted to select the necessary and sufficient components of the topological vector to distinguish at least three kinds of information: the type of the central atom, the chemical short-range ordering (CSRO, \cite{Cowley1950}), and the structural information (e.g. B2 and L1\textsubscript{2}). The CSRO parameter $\alpha_1$ is computed for each structure on the first shell of neighbors by 	
\begin{equation}
	\alpha_{1}=1-\frac{\overline{\nu_{\beta}}}{\nu.\overline{c_{\beta}}}
\end{equation}
where $\beta$ is the type of the central atom,  $\overline{\nu_{\beta}}$ is the number of neighbors of a type different from $\beta$, $\nu$ is the total number of neighbors and $\overline{c_{\beta}}$ is the concentration of atoms of a type different from $\beta$ in the original crystalline configuration from which the structure comes from.

First, to check that the chemical oriented distance vector is able to retrieve the chemical information, we use a classification Random Forest (implemented in the Python package \texttt{scikit-learn} \cite{Pedregosa2011})  with 100 decision trees. We performed two classification tasks: one for the type of the central atom, and one related to the CSRO. As only 28 different values were obtained for the CSRO on the local noisy crystalline structures (16 values for structures centered in Ni and 12 values for structures centered in Al), with more than 1000 structures in each case, we consider those values as different classes. For each classification task, we consider 100 stratified  train test splits with proportions 90\%/10\%. All the data are use for predicting the type of the central atom, whereas the CSRO is differentiate by the central atom. Table~\ref{Tab:RFPCCSRO} gives the accuracy, recall and precision of the predictions made on each test set. The high values of these scores are a good indication that the chemical oriented distances vector is a proper representative of the information in terms of central atom and CSRO. 
	
\begin{table}[t]
	\centering
	\begin{tabular}{lccc}
		\hline
		\hline
		& Type of central atom & \multicolumn{2}{c}{CSRO}  \\
		& \makebox[3cm][c]{} & \makebox[3cm][c]{Central atom Ni} & \makebox[3cm][c]{Central atom Al}\\
		\hline
		Accuracy & $1.00 \pm 0.00$ & $0.99 \pm 0.01$ & $0.98 \pm 0.01$ \\
		Recall & $1.00 \pm 0.00$ & $0.99 \pm 0.01$ & $0.99 \pm 0.01$ \\
		Precision & $1.00 \pm 0.00$ & $0.99 \pm 0.01$ & $0.99 \pm 0.01$ \\
		\hline
	\end{tabular}	
	\caption{Means and standard deviations of the accuracy, recall and precision metrics of the random forest over 100 train/test splits.}
	\label{Tab:RFPCCSRO}
\end{table}

\begin{table}[t]
	\begin{tabular}{lcccccc}
		\hline
		\hline
		& \multicolumn{6}{c}{Structural information}  \\
		& \multicolumn{3}{c}{Central atom Ni} & \multicolumn{3}{c}{Central atom Al}\\
		& \makebox[2cm][c]{$\text{TV}_0$} & \makebox[2cm][c]{$\text{TV}_1$}& \makebox[2cm][c]{$\text{TV}_2$} & \makebox[2cm][c]{$\text{TV}_0$} &
		\makebox[2cm][c]{$\text{TV}_1$}& \makebox[2cm][c]{$\text{TV}_2$}\\
		Dimension & 156 & 86 & 19 & 156 & 86 & 19\\
		\hline
		Accuracy & $0.98 \pm 0.01$ & $0.99 \pm 0.01$ & $0.45 \pm 0.01$ & $0.94 \pm 0.01$ & $0.99 \pm 0.01$ & $0.47 \pm 0.01$ \\
		Recall & $0.99 \pm 0.01$ & $0.99 \pm 0.01$ & $0.43 \pm 0.01$ & $0.94 \pm 0.01$ & $0.99 \pm 0.01$ & $0.41 \pm 0.01$ \\
		Precision & $0.99 \pm 0.01$ & $0.99 \pm 0.01$ & $0.51 \pm 0.02$ & $0.94 \pm 0.01$ & $0.99 \pm 0.01$ & $0.39 \pm 0.01$ \\
		\hline
	\end{tabular}	
	\caption{Means and standard deviations of the accuracy, recall and precision metrics computed from the predictions on test sets made by 100 different random forest classifiers trained with the different constituent parts of the topological vectors (TV) on labels representing structures with two shells of neighbors for each subset of structures with a central Ni or Al atom. $\text{TV}_0$ stands for the 0-dimensional part of the topological vectors: $\text{TV}^{H^{\text{NiNi}}_{0}}$, $\text{TV}^{H^{\text{NiAl}}_{0}}$ and $\text{TV}^{H^{\text{AlAl}}_{0}}$ ;
		$\text{TV}_1$ corresponds the 1-dimensional part of the topological vectors: $\text{TV}^{H^{\text{NiNi}}_{1}}$ and $\text{TV}^{H^{\text{AlAl}}_{1}}$ \footnote{As mentioned above $H_1^{\text{NiAl}}$, with only points at infinity in the PD, was not considered because it increases drastically the dimension of the vector.} ; and $\text{TV}_2$ gives the 2-dimensional part of the topological vectors:  $\text{TV}^{H^{\text{NiNi}}_{2}}$ and $\text{TV}^{H^{\text{AlAl}}_{2}}$.}
	\label{Tab:RFCNA}
\end{table}

For the topological vectors, we applied the same supervised learning process. We set a label for each distinct CNA structural signature. This leads to a total number of 33 labels (17 labels for structures with a central atom Ni and 16 labels for structures with a central atom Al).
Table~\ref{Tab:RFCNA} gives the means and standard deviations of the accuracy, recall and precision metrics computed on the predictions made on each test set, through the hundred random forest models.	

Based on these scores and to get a moderate dimension for the descriptor (according to our training set), we choose to keep the topological vectors constructed from  $H_{1}^{AA}$ and $H_{1}^{BB}$ to encode the structural information of our local structure.

\section*{Additional tables and figures}

We give here additional tables and figures that support the main text.

\begin{table}[H]
	\begin{tabular}{lrrrrrr}
		\hline\hline
		(\%) & \makebox[2cm][r]{0.3 ns} & \makebox[2cm][r]{0.5 ns} & \makebox[2cm][r]{0.7 ns} & \makebox[2cm][r]{0.9 ns} & \makebox[2cm][r]{2 ns} & \makebox[2cm][r]{5 ns}\\
		\hline
		B2\textsuperscript{Al} & $0.008^{1}$ & $0.236^{1}$ & $1.731^{1}$ & $6.018^{1}$ & $15.576^{2}$ & $18.721^{2}$ \\
		B2\textsuperscript{Ni} & $0.016^{3}$ & $0.245^{3}$ & $1.758^{3}$ & $6.025^{3}$ & $15.626^{4}$ & $18.701^{4}$ \\
		BCC\textsuperscript{Al}\textsubscript{$\alpha_1 \notin \{-0.1429, 0\}$} & $0.003^{2}$ & $0.104^{2}$ & $1.082^{3}$ & $3.950^{3}$ & $7.875^{2}$ & $7.579^{2}$ \\
		BCC\textsuperscript{Ni}\textsubscript{$\alpha_1 \notin \{-0.1429, 0\}$} & $0.001^{1}$ & $0.084^{5}$ & $0.843^{5}$ & $3.102^{5}$ & $5.817^{4}$ & $5.480^{4}$ \\
		BCC\textsuperscript{Al}\textsubscript{$\alpha_{1} = 0$} & $0.004^{2}$ & $0.086^{2}$ & $0.851^{2}$ & $3.004^{3}$ & $5.602^{3}$ & $5.185^{3}$ \\
		BCC\textsuperscript{Ni}\textsubscript{$\alpha_{1} = 0$} & $0.007^{3}$ & $0.132^{3}$ & $1.197^{3}$ & $4.400^{3}$ & $9.158^{2}$ & $8.949^{2}$ \\
		\hline\hline
	\end{tabular}	
	\caption{Evolution of B2 ($\alpha_{1} = -0.1429$), BCC chemically ordered ($\alpha_1 \notin \{-0.1429, 0\}$) and BCC chemically disordered ($\alpha_{1} = 0$) crystalline structures in Al\textsubscript{50}Ni\textsubscript{50} during nucleation for central Al and Ni atoms. The superscripts correspond to the number of clusters in which the structures are distributed.}
	\label{Tab:Al50Ni50_evolution}
\end{table}

\begin{figure}[H]
	\centering
	\includegraphics[width=1.0\textwidth]{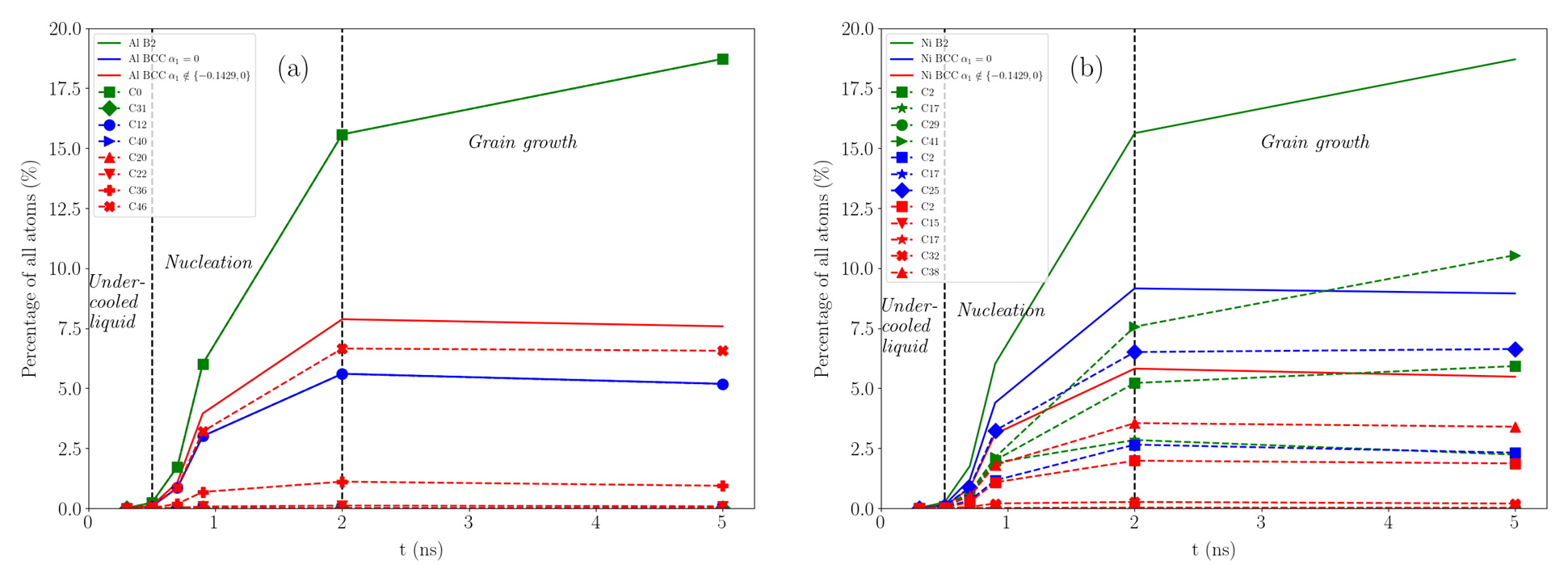}
	\caption{Total evolution of B2 (green), BCC chemically disordered (blue) and BCC chemically ordered (red) in Al\textsubscript{50}Ni\textsubscript{50} during nucleation for central (a) Al and (b) Ni atoms. We also display the evolution of the corresponding clusters.}
\end{figure}

\begin{figure}[H]
	\centering
	\includegraphics[width=1.0\textwidth]{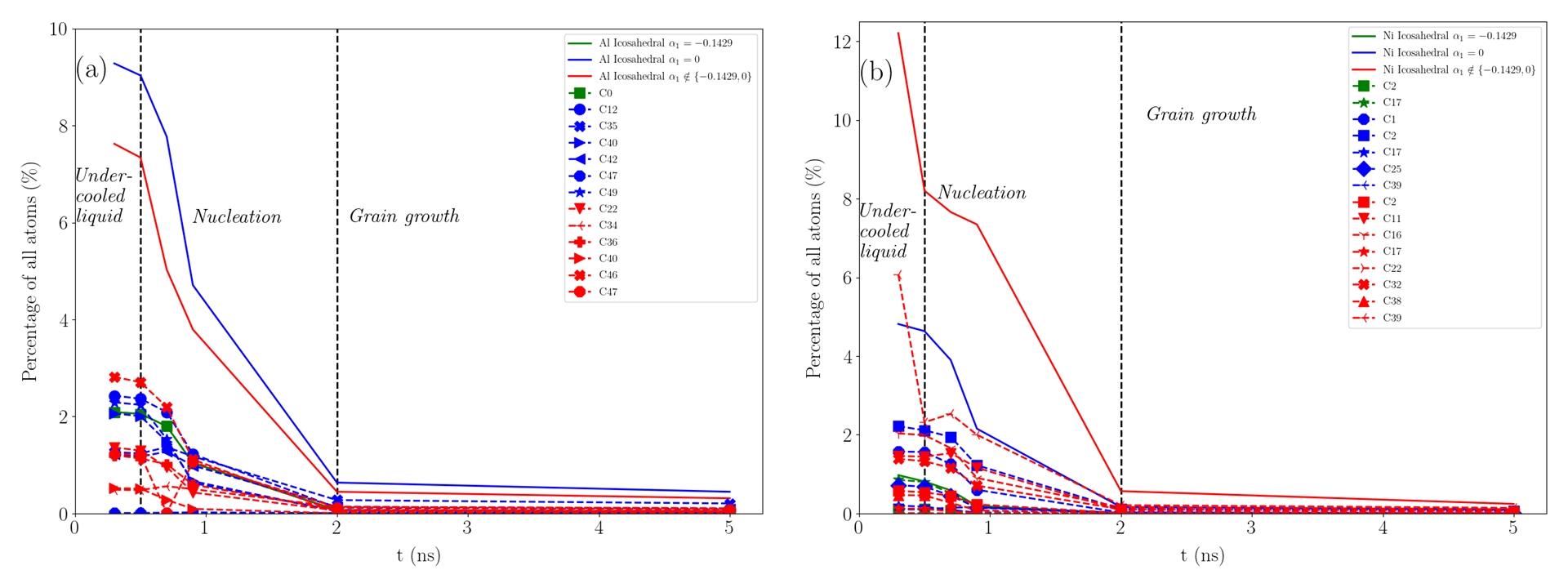}
	\caption{Total evolution of five-fold symmetry (icosahedral), with a B2 ordering (green), chemically disordered (blue) and chemically ordered (red) in Al\textsubscript{50}Ni\textsubscript{50} during nucleation for central (a) Al and (b) Ni atoms. We also display the evolution of the corresponding clusters.}
\end{figure}

\begin{table}[H]
	\begin{tabular}{lrrrrrr}
		\hline\hline
		(\%) & \makebox[2cm][r]{9.6 ns} & \makebox[2cm][r]{10 ns} & \makebox[2cm][r]{10.25 ns} & \makebox[2cm][r]{10.50 ns} & \makebox[2cm][r]{11 ns} & \makebox[2cm][r]{14 ns}\\
		\hline
		BCC\textsuperscript{Al}\textsubscript{$\alpha_{1} = -0.333$} & $0.001^{2}$ & $0.006^{2}$ & $0.027^{2}$ & $0.029^{2}$ & $0.018^{2}$ & $0.019^{2}$ \\
		BCC\textsuperscript{Ni}\textsubscript{$\alpha_{1} = -0.333$} & $0$ & $0$ & $0$ & $0$ & $0$ & $0$ \\
		BCC\textsuperscript{Al}\textsubscript{$\alpha_1 \notin \{-0.333, 0\}$} & $0.005^{4}$ & $0.130^{5}$ & $0.460^{5}$ & $0.454^{4}$ & $0.301^{5}$ & $0.297^{5}$ \\
		BCC\textsuperscript{Ni}\textsubscript{$\alpha_1 \notin \{-0.333, 0\}$} & $0.024^{4}$ & $0.620^{7}$ & $2.259^{8}$ & $2.734^{8}$ & $0.385^{5}$ & $0.414^{5}$ \\
		BCC\textsuperscript{Al}\textsubscript{$\alpha_{1} = 0$} & $0.006^{1}$ & $0.096^{1}$ & $0.322^{1}$ & $0.422^{1}$ & $0.229^{1}$ & $0.230^{1}$ \\
		BCC\textsuperscript{Ni}\textsubscript{$\alpha_{1} = 0$} & $0.002^{2}$ & $0.002^{2}$ & $0.002^{2}$ & $0$ & $0$ & $0$ \\
		\hline
		HCP\textsuperscript{Al}\textsubscript{$\alpha_{1} = -0.333$} & $0$ & $0$ & $0.005^{4}$ & $0.160^{4}$ & $0.557^{4}$ & $0.625^{4}$ \\
		HCP\textsuperscript{Ni}\textsubscript{$\alpha_{1} = -0.333$} & $0.002^{2}$ & $0.004^{3}$ & $0.051^{5}$ & $1.349^{5}$ & $5.248^{5}$ & $5.793^{5}$ \\
		HCP\textsuperscript{Al}\textsubscript{$\alpha_1 \notin \{-0.333, 0\}$} & $0.002^{2}$ & $0.004^{3}$ & $0.086^{4}$ & $0.676^{4}$ & $2.427^{5}$ & $2.618^{5}$ \\
		HCP\textsuperscript{Ni}\textsubscript{$\alpha_1 \notin \{-0.333, 0\}$} & $0.004^{4}$ & $0.027^{11}$ & $0.422^{12}$ & $2.263^{13}$ & $7.123^{11}$ & $7.699^{11}$ \\
		HCP\textsuperscript{Al}\textsubscript{$\alpha_{1} = 0$} & $0$ & $0.002^{2}$ & $0.03^{3}$ & $0.235^{2}$ & $0.822^{2}$ & $0.860^{2}$ \\
		HCP\textsuperscript{Ni}\textsubscript{$\alpha_{1} = 0$} & $0.003^{3}$ & $0.008^{4}$ & $0.152^{5}$ & $1.739^{5}$ & $5.989^{5}$ & $6.489^{5}$ \\
		\hline
		L1\textsubscript{2}\textsuperscript{Al} & $0$ & $0$ & $0.008^{4}$ & $0.122^{4}$ & $0.549^{4}$ & $0.653^{4}$ \\
		L1\textsubscript{2}\textsuperscript{Ni} & $0$ & $0.002^{2}$ & $0.033^{5}$ & $1.090^{5}$ & $5.243^{6}$ & $6.083^{6}$ \\
		FCC\textsuperscript{Al}\textsubscript{$\alpha_1 \notin \{-0.333, 0\}$} & $0$ & $0.004^{4}$ & $0.065^{4}$ & $0.564^{5}$ & $2.558^{6}$ & $2.858^{6}$ \\
		FCC\textsuperscript{Ni}\textsubscript{$\alpha_1 \notin \{-0.333, 0\}$} & $0.002^{1}$ & $0.014^{8}$ & $0.320^{12}$ & $1.935^{11}$ & $7.346^{12}$ & $8.305^{12}$ \\
		FCC\textsuperscript{Al}\textsubscript{$\alpha_{1} = 0$} & $0$ & $0$ & $0.0190^{2}$ & $0.190^{2}$ & $0.846^{2}$ & $0.931^{3}$ \\
		FCC\textsuperscript{Ni}\textsubscript{$\alpha_{1} = 0$} & $0$ & $0.005^{5}$ & $0.116^{5}$ & $1.441^{5}$ & $6.08^{5}$ & $6.948^{5}$ \\
		\hline\hline
	\end{tabular}	
	\caption{Evolution of the BCC, HCP, and FCC crystalline structures with a L1\textsubscript{2} ordering $\alpha_{1} = -0.333$, chemically ordering $ \alpha_1 \notin \{-0.333, 0\}$, and chemically disordered ($\alpha_{1} = 0$) in Al\textsubscript{25}Ni\textsubscript{75} during nucleation for central Al and Ni atoms. The superscripts correspond to the number of clusters in which the structures are distributed.}
	\label{Tab:Al25Ni75_evolution}
\end{table}

\begin{figure}[H]
	\centering
	\includegraphics[width=1.0\textwidth]{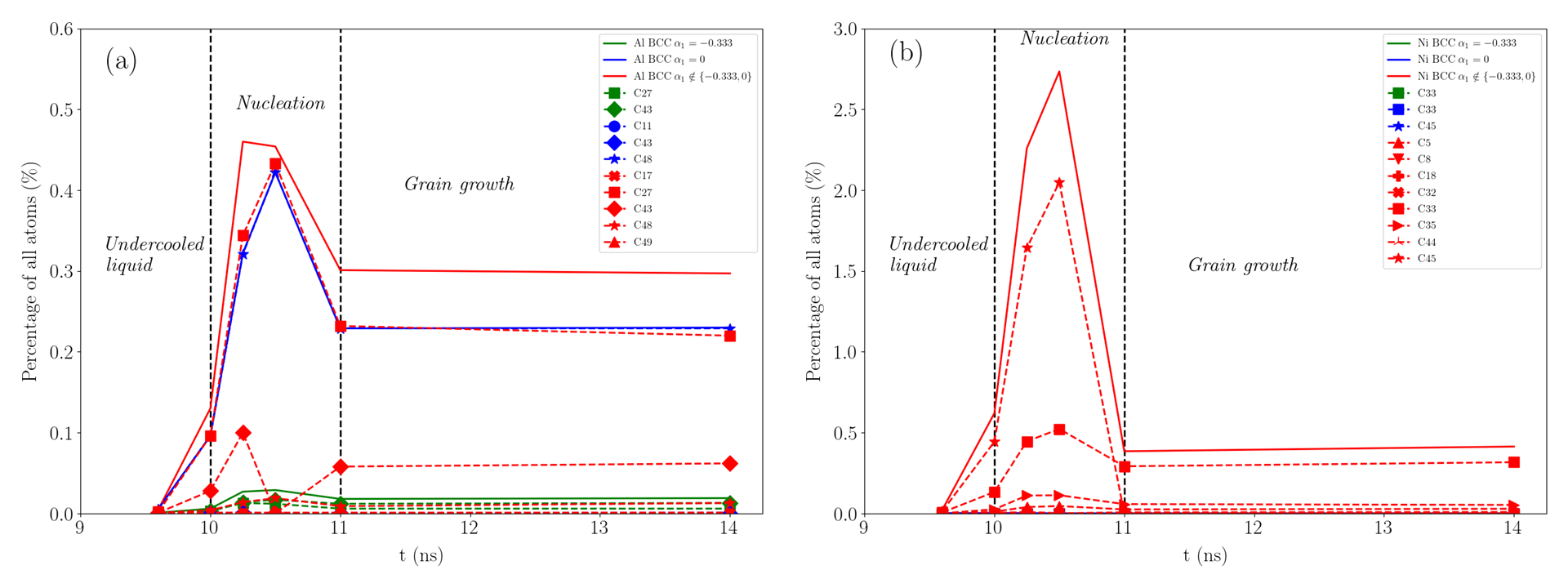}
	\caption{Total evolution of BCC with a L1\textsubscript{2} ordering (green), BCC chemically disordered (blue) and BCC chemically ordered (red) in Al\textsubscript{25}Ni\textsubscript{75} during nucleation for central (a) Al and (b) Ni atoms. We also display the evolution of the corresponding clusters.}
\end{figure}

\begin{figure}[H]
	\centering
	\includegraphics[width=1.0\textwidth]{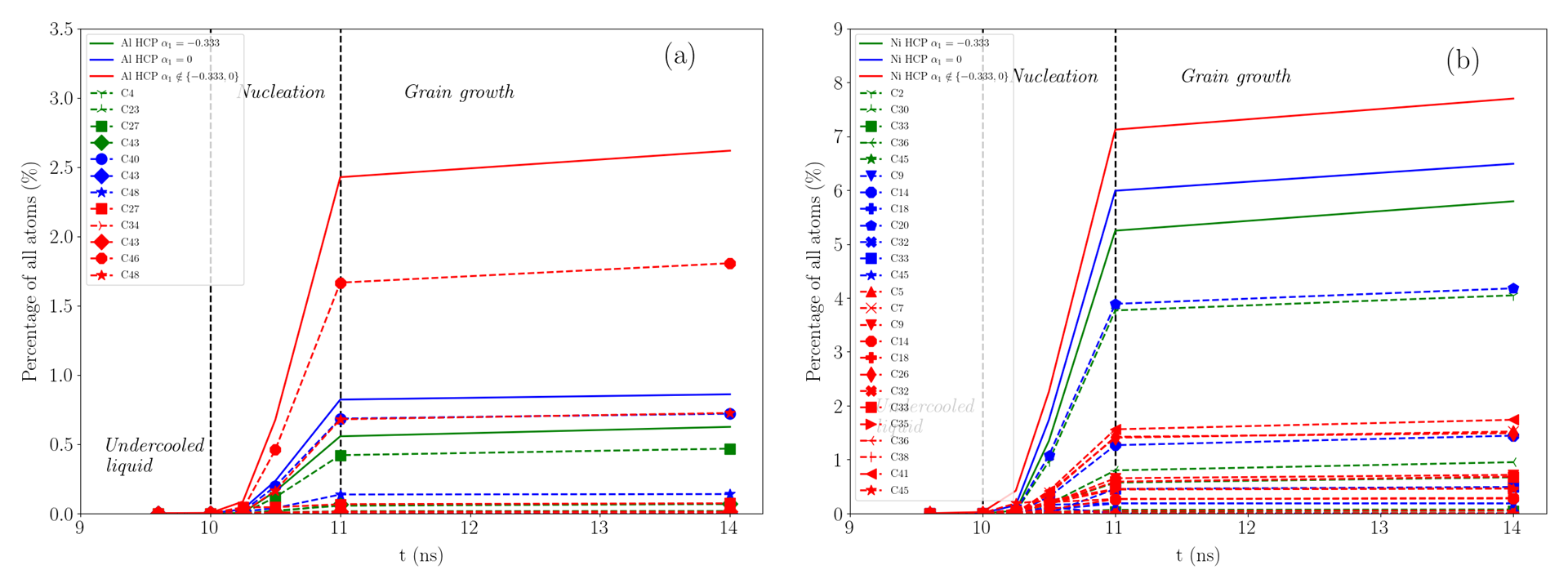}
	\caption{Total evolution of HCP with a L1\textsubscript{2} ordering (green), HCP chemically disordered (blue) and HCP chemically ordered (red) in Al\textsubscript{25}Ni\textsubscript{75} during nucleation for central (a) Al and (b) Ni atoms. We also display the evolution of the corresponding clusters.}
\end{figure}

\begin{figure}[H]
	\centering
	\includegraphics[width=1.0\textwidth]{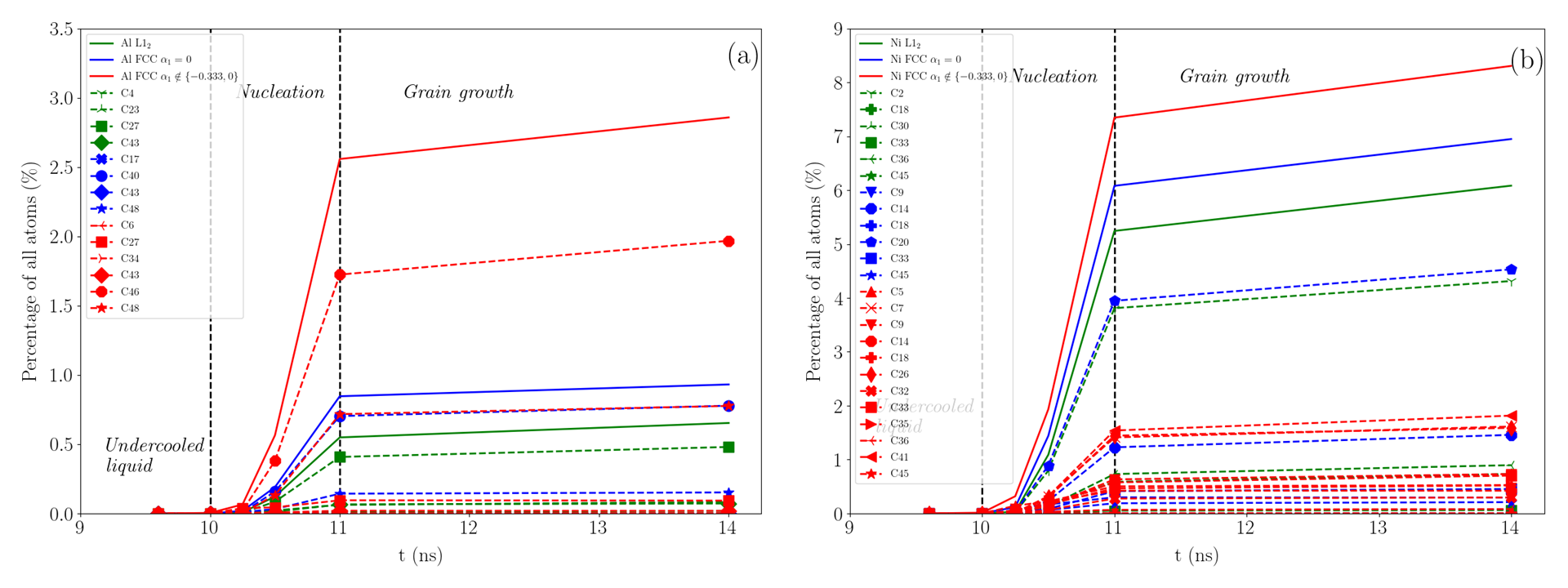}
	\caption{Total evolution of L1\textsubscript{2} (green), FCC chemically disordered (blue) and FCC chemically ordered (red) in Al\textsubscript{25}Ni\textsubscript{75} during nucleation for central (a) Al and (b) Ni atoms. We also display the evolution of the corresponding clusters.}
\end{figure}

\begin{figure}[H]
	\centering
	\includegraphics[width=1.0\textwidth]{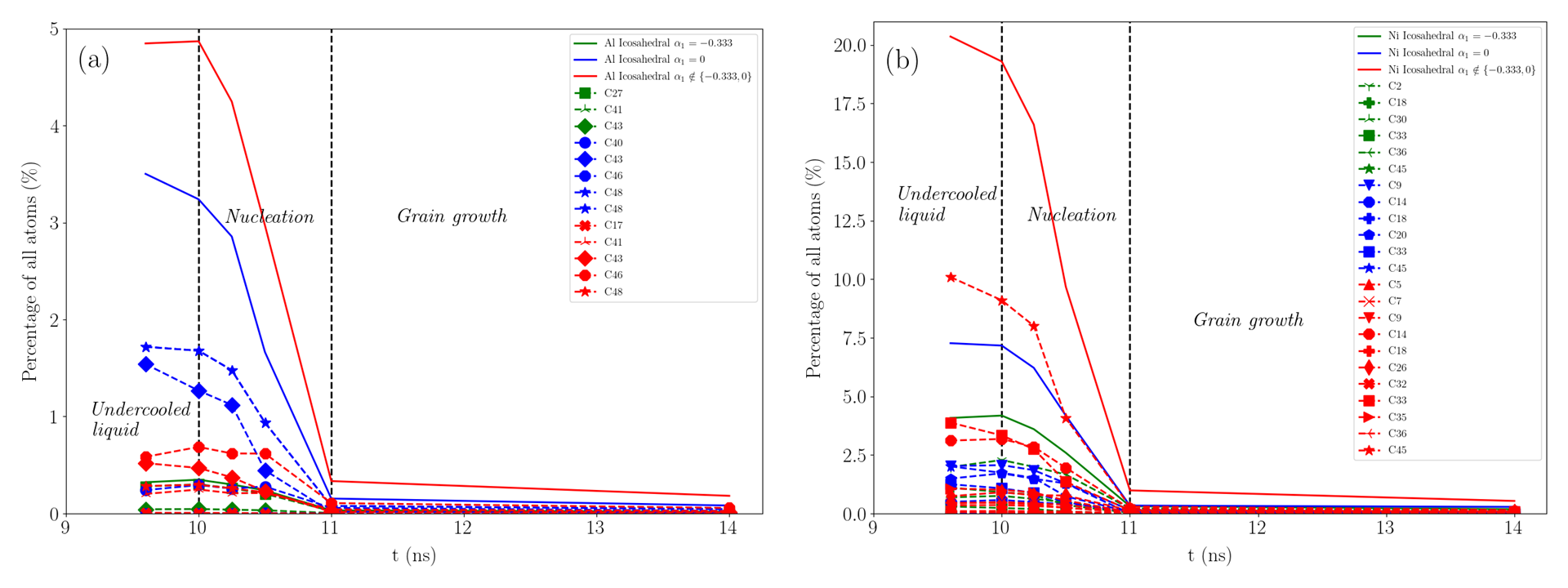}
	\caption{Total evolution of five-fold symmetry (icosahedral), with a L1\textsubscript{2} ordering (green), chemically disordered (blue) and chemically ordered (red) in Al\textsubscript{25}Ni\textsubscript{75} during nucleation for central (a) Al and (b) Ni atoms. We also display the evolution of the corresponding clusters.}
\end{figure}

\end{document}